\documentclass[10pt]{article}
\usepackage{epsfig}
\usepackage{chicago}
\usepackage{authblk}
\usepackage{color}
\usepackage{amsmath}
\usepackage{amssymb}
\begin{document}
\title{Electrostatic activation of prebiotic chemistry in substellar atmospheres}
\author[1]{C. R. Stark\thanks{email:~craig.stark@st-andrews.ac.uk; tel:~01334 463053; fax: 01334 463104}}
\author[1]{Ch. Helling}
\author[2]{D. A. Diver}
\author[1]{P. B. Rimmer}
\affil[1]{SUPA, School of Physics and Astronomy, University of St Andrews, St Andrews, KY16 9SS, UK}
\affil[2]{SUPA, School of Physics and Astronomy, University of Glasgow, Glasgow, G12 8QQ, UK}
\date{Keywords: Dust, Exoplanets, Prebiotic Chemistry, Plasmas. \\
Running title: Electrostatic activation}
\maketitle{}
\begin{abstract}
Charged dust grains in the atmospheres of exoplanets may play a key role in the formation of prebiotic molecules, necessary to the origin of life.  Dust grains submerged in an atmospheric plasma become negatively charged and attract a flux of ions that are accelerated from the plasma.  The energy of the ions upon reaching the grain surface may be sufficient to overcome the activation energy of particular chemical reactions that would be unattainable via ion and neutral bombardment from classical, thermal excitation.  As a result, prebiotic molecules or their precursors could be synthesised on the surface of dust grains that form clouds in exoplanetary atmospheres.  This paper investigates the energization of the plasma ions, and the dependence on the plasma electron temperature, in the atmospheres of substellar objects such as gas giant planets.  Calculations show that modest electron temperatures of $\approx 1$~eV ($\approx 10^{4}$~K) are enough to accelerate ions to sufficient energies that exceed the activation energies required for the formation of formaldehyde, ammonia, hydrogen cyanide and the amino acid glycine.
\end{abstract}
\section{Introduction}
Dust is ubiquitous in the Universe, growing in a variety of diverse environments from the interstellar medium; the Earth's troposphere; to the atmospheres of gas giant exoplanets, where mineral dust clouds form.  Without it, the universe would lack the building blocks essential for planetary formation, and possibly even for the synthesis of life itself.  Mineral dust clouds play an important role in substellar atmospheres (objects whose mass is sufficiently low that they cannot sustain hydrogen fusion).  Their formation causes a significant depletion of the ambient gas, which alters the subsequent gas-phase chemistry and the consequent observable absorption features~\cite{bilger2013}.  Notably, oxygen is depleted (bringing the C/O ratio closer to one) and so the grain surface may find itself in an environment more favourable to carbon-bonding molecules (hence promoting organic chemistry) more than a dust free atmosphere.  The dust grains that compose the clouds present a rich catalytic surface conducive to the formation of complex, prebiotic molecules sourced from the surrounding environment~\cite{hill2003,charnley2001}.  In substellar atmospheres, dust forms at a certain altitude, depleting the local gas phase species.  The dust particles gravitationally settle, falling to lower altitudes where they evaporate and are mixed by convective processes which replenish the atmospheric gas~\cite{helling2008,tsuji2002,allard2001,burrows1997,marley2002,morley2012}.  

If regions of the atmospheric gas are ionised the dust will become charged and the surrounding environment becomes electrically activated, allowing otherwise improbable chemical reactions.  Miller and Urey were among the first to demonstrate the importance of electrical activation in the synthesis of prebiotic molecules~\cite{miller1953,miller1959}.  They considered a planetary atmosphere rich in H$_{2}$, CH$_{4}$, NH$_{3}$ and H$_{2}$O and successfully synthesised prebiotic amino acids (e.g. Glycine, Alanine, etc) and other biological molecules (e.g. Urea, Lactic acid, etc) when such a gas mixture participates in an electrical discharge.  Amino acids are critical ingredients for life on Earth since they are required for the formation of proteins, peptides and enzymes.  A similar yield of amino acids was also found in experiments carried out to investigate prebiotic synthesis in steam-rich volcanic eruptions~\cite{johnson2008}.  Nebular lightning has also been postulated as a mechanism for the processing of nebular dust in order to explain the oxygen isotopic content in the solar system~\cite{nuth2012}.  

It is important to note that the atmospheric conditions simulated in the Miller-Urey experiment do not correspond to those thought to exist in the primitive Earth since CH$_{4}$ and NH$_{3}$ were not believed to be very abundant.  Modern volcanic gas emission show that most of the carbon and nitrogen exists as CO$_{2}$ and N$_{2}$ respectively~\cite{kasting1993};  if past volcanic emission was similar, this would infer a lower abundance of CH$_{4}$ and NH$_{3}$ than that used in the Miller-Urey experiment.  For atmospheres more representative of primitive Earth, no significant organic molecules are produced using electrical (sparking) discharges~\cite{kasting2003}.  However, the presence of hydroxy acids in the famous Murchison meteorite indicate that the Strecker mechanism (triggered by a Miller-Urey-type electrical discharge) may be responsible for the extraterrestrial synthesis of amino acids~\cite{peltzer1978}.

Laboratory experiments have established that irradiation of interstellar ice analogues with polarised and unpolarised UV radiation leads to the formation of amino acids~\cite{bernstein2002,nuevo2006,nuevo2007,caro2002,sorrell2001,woon2002,elsila2007}.  Furthermore, Nuevo et al., 2006, 2007 found that the enantiomeric excess in the ice mixtures (H$_{2}$, CO, CO$_{2}$, CH$_{4}$, CH$_{3}$OH, NH$_{3}$) as a result of the circularly polarised radiation was very small.  The processes leading to the formation of amino acids has long been an outstanding problem.
Elsila et al., 2007 experimentally tested if the formation of amino acids in interstellar ices occurred via Strecker-type synthesis~\cite{bernstein2002} or radical-radical interactions~\cite{woon2002,sorrell2001} finding that neither satisfactorily explained their laboratory results.

There are collective consequences if grains are charged.  Inter-grain electrical discharges (sparking)~\cite{helling2011a} is a collective effect that activates the ambient gas, creating physical conditions similar to those thought responsible for life in early planetary atmospheres, analogous to the Miller-Urey experiment~\cite{miller1953,miller1959}.  This process is amplified by the very strong electric fields that develop at the poles of prolate grains~\cite{stark2006}.  Moreover, prolate grains have two key advantages over spherical grains: a greater surface area which promotes surface catalysis; and inhomogeneous electric fields. In a magnetised medium, such grains become aligned to the ambient field, polarising passing electromagnetic radiation~\cite{davis1949} and altering the subsequent chirality-dependent biological chemistry~\cite{bailey1998}. Polarization also provides an extremely useful diagnostic to probe the ambient environment.

Classically, in the gas-phase, grain surface chemistry occurs via the absorption of neutral species via the stochastic kinetic motions of the gas.  The probability of a particular chemical reaction occurring depends upon the reaction rate coefficients which itself is determined by the underlying velocity distribution of the participating neutral species.  A particular reaction will occur provided that the thermal energy of the ambient gas can surmount the required activation energy.  However, in an atmospheric dusty plasma  (a plasma containing dust particles) the absorption of species is electrostatically driven and the resulting reactivity on the grain surface can be amplified.  

Consider a dusty plasma in the atmosphere of a substellar object such as a giant gas exoplanet.  The dust particles will be negatively charged and as a result a plasma sheath (an electron depleted region, see Section~\ref{es}) forms around the particle.  As a consequence, the ionic flux at the grain surface increases as the plasma ions are attracted to and are accelerated towards the grain surface.  Upon reaching the surface the ions have fallen through an electrostatic potential and have been energised.  In comparison to the neutral case, the ionic flux is enhanced and the ionic energy amplified, increasing the probability that chemical reactions will occur and that reactions with higher activation energies are accessible.  In this way, charged particle surfaces help catalyse chemical reactions otherwise inaccessible at such low-temperatures present in planetary atmospheres.

The aim of this paper is to investigate the energization of plasma ions as they are accelerated from the bulk plasma to the surface of a charged dust grain.  If the ions gain a significant amount of energy, this energy may be sufficient to overcome the activation barrier of certain chemical reactions that would be inaccessible via classical thermal ion and neutral fluxes.  This may lead to an increase in the creation of prebiotic molecules on the grain surface.  This paper is structured as follows: Section~\ref{plasmas} discusses the sources of ionization and the generation of plasmas in substellar atmospheres, focusing on Alfv\'{e}n ionisation and summarising the findings of Stark and co-workers~\cite{stark2013};  Section~\ref{es} discusses the charging of dust grains immersed in a plasma and how the plasma ions can be energised by the local electric fields of the dust grains; and Section~\ref{discuss} discusses the electrostatic activation of particular chemical reactions as a result. We are particularly interested in these plasma processes in the atmospheres of exoplanets.  We will use an example atmosphere using the \textsc{Drift-Phoenix} model atmosphere and cloud formation code, characterized by $\log{g}=3.0$~and~$T_{\rm eff}=1500$~K.  In such an atmosphere the gas-phase temperature varies from $T_{\rm gas}\approx 500$~K at low atmospheric pressures to $T_{\rm gas}\approx3300$~K at high atmospheric pressures (see Figure~\ref{fig1}).  \textsc{Drift-Phoenix} considers collisionally dominated atmospheres where the gas-phase and dust cloud particles have the same temperature $T_{\rm dust}=T_{\rm gas}$; therefore, for the dust populated regions of the atmosphere the dust temperature is $T_{\rm dust}\approx600-1900$~K.

\section{Substellar atmospheric plasmas\label{plasmas}}
Electromagnetic emission from substellar objects in the radio and X-ray frequency bands infers the presence of plasmas in their atmospheres.   A number of ionization processes occur in substellar atmospheres that produce volumes of gas-plasma mixtures.  For an ionised gas to exhibit plasma behaviour, the dynamics of the ionized particles must be dominated by long-range collective electromagnetic effects and not by short-range collisions (such as in a neutral gas)~\cite{chen1984}.  An ionised gas will become a plasma when the degree of ionization is $\geq 10^{-7}$~\cite{diver2001,fridman2008}.  On its own, thermal ionization is insufficient to ionize a significant fraction of the neutral substellar atmosphere so that the resulting ionised gas is a plasma~\cite{isabel2013}.  

In substellar atmospheres where cloud formation occurs, gas discharge events (i.e. lightning) involving charged cloud particles can generate a local plasma volume with sufficiently high number densities~\cite{helling2011a}.  Helling and co-workers have investigated the conditions for inter-grain and large-scale electrical discharge events to occur in the atmospheres of substellar objects~\cite{helling2013,bailey2013}.  Cosmic-ray interactions with the atmosphere can also ionize the atmospheric gas~\cite{rimmer2013}.  In addition, turbulence-induced dust-dust collisions can ionise fractions of the atmosphere, enhancing the local degree of ionization~\cite{helling2011b}.  Alfv\'{e}n ionization is an alternative process by which volumes of atmospheric plasma can be efficiently generated.  The following is a summary of the Alfv\'{e}n ionisation process in substellar atmospheres~\cite{stark2013}.

In Alfv\'{e}n ionization a flow of neutral gas impinges on a low-density, magnetised seed plasma.  The kinetic energy of the neutral flow is $\frac{1}{2}m_{\rm gas}v_{0}^{2}$; where, $m_{\rm gas}$ is the mass of a neutral atom or molecule and $v_{0}$ is the flow speed.  By magnetised we mean that the electrons are localised and their motion is impeded in the perpendicular direction to the ambient magnetic field due to the Lorentz force.  The neutral atoms of the flow collide with and elastically scatter the plasma ions, sending them off to participate in Larmor orbits due to the magnetic field.  As more and more ions are displaced, an unscreened pocket of electrons, or charge imbalance, is exposed.  The electrons are unable to rectify the charge imbalance due to their magnetically restricted motion.  The charge imbalance stops growing when further ionic displacement is inhibited by the local electric field of the exposed electrons.  This occurs when the electrostatic potential energy of the charge imbalance equals the kinetic energy of the initial neutral flow.  At this point, the pocket of electrons due to their mutual self-electrostatic repulsion, are accelerated parallel to the magnetic field to an energy equal to $\frac{1}{2}m_{\rm gas}v_{0}^{2}$.  If these electrons go on to collide with the surrounding atmospheric neutral gas and their kinetic energy equals or exceeds the electrostatic potential energy required to ionise the neutral, more electrons and ions are liberated.  

Alfv\'{e}n ionization requires: a low-density, magnetized seed plasma; and a neutral gas flow that has a kinetic energy that exceeds the electrostatic potential energy required to ionise a neutral atom or molecule of the surrounding atmospheric gas:  $\frac{1}{2}m_{\rm gas}v_{0}^{2}\geq e\phi_{I}$.  The critical flow speed required is $v_{c}=(2e\phi_{I}/m_{\rm gas})^{1/2}$, where $\phi_{I}$ is the first ionization potential of the species of interest.  In substellar atmospheres the seed plasma can be generated from local electrical discharge events (sparking) in mineral clouds and cosmic-ray bombardment of the atmosphere, enhancing the ambient electron number density by $n_{e}\approx 10^{23}$~m$^{-3}$~\cite{uman1964,guo2009,chang2010} and~$n_{e}\approx10^{10}$~m$^{-3}$~\cite{rimmer2013} respectively.  

The low-density, seed plasma must also be magnetized.  Giant gas planets and Brown Dwarfs have typical large-scale magnetic flux densities estimated to be of the order $10$~G and $1$~kG respectively~\cite{donati2009,reiners2012,christensen2009,sanchez2004,shulyak2011}, which are sufficient to ensure that the seed plasma is magnetized for a large fraction of the atmosphere.  In general, for the neutral species expected to populate the envelopes of substellar objects, the critical neutral gas flow speeds required are $\approx O(1-10~\textnormal{km~s}^{-1})$.  Studies of substellar atmospheric circulation, flows and winds have shown that flow speeds of $v_{0}\approx 1-10$~km~s$^{-1}$ ~\cite{showman2002,showman2008,showman2009,menou2009,rauscher2010,cooper2005,dobbs2008,dobbs2010,dixon2012,lewis2010,heng2011}~(and possibly higher) are attainable.  These flow speeds are averaged over an underlying particle distribution of speeds, and so a population of high-energy particles will be present that will be able to reach larger critical speeds.

If the required criteria can be met, Alfv\'{e}n ionization is an efficient process by which volumes of plasma with degrees of ionization ranging from $10^{-6}-1$ can be obtained.  The atmospheric plasma volumes can vary from small to large scales.  If we assume that in a localized atmospheric volume the entirety of a particular target species can be 100\% ionised, then the gas-phase species that has the greatest relative density will yield the greatest degree of ionisation resulting from Alfv\'{e}n ionization.  In the model substellar atmospheres considered He, Fe, Mg, Na, H$_{2}$, CO, H$_{2}$O, N$_{2}$ and SiO for all atmospheric pressures yield a degree of ionization $\geq 10^{-7}$, producing an atmospheric plasma.  In reality, the resulting plasma volumes will be composed of multi-species plasmas (composing of NH$_{3}$, CH$_{4}$, CO, etc), even containing ionic species that if ionized on their own would have an insufficient degree of ionization to constitute a plasma.  Of particular interest in the synthesis of prebiotic molecules is NH$_{3}$ and CH$_{4}$ which can be dominant molecular species in substellar atmospheres~\cite{helling2008b,bilger2013}

The properties and characteristics of the plasma, such as the plasma temperature, depend on the process which created it.  Since the electrons are much less massive than the neutrals and ions, they only lose a small fraction of their energy following a collision.  As a result, the plasma electrons and ions will not necessarily be in thermal equilibrium with each other or the ambient gas and the electron temperature can differ from the ion and gas-phase temperature.   Typical terrestrial, gas discharge electron temperatures are $T_{e}\approx1-100$~eV ($\approx10^{4}-10^{6}$~K)~\cite{fridman2008,diver2001}; in substellar atmospheres electrons from thermal ionization are in thermal equilibrium with the ambient neutral gas ($T_{e}\approx T_{\rm gas}$) and have electron temperatures of $T_{e}\approx O(0.01-0.1$~eV$)$ ($\approx O(10^{2}-10^{3}$~K$)$). 

In this paper, we shall assume an atmosphere that contains plasma volumes composed of multi-species plasmas (e.g. Fe, H$_{2}$, NH$_{3}$, CH$_{4}$, CO, H$_{2}$O, SiO, etc) populated with dust grains, with plasma electron temperatures ranging from $T_{e}\approx 0.01-100$~eV ($\approx O(10^{2}-10^{3}$~K$)$).

\section{Electrostatic activation of prebiotic chemistry\label{es}}
Alfv\'{e}n ionization creates volumes of plasma in the atmospheres of substellar objects, which are populated with dust cloud particles.  Let's consider a dust grain immersed in an electron-ion (a multi-species) plasma that is in thermal equilibrium ($T_{e}=T_{i}=T_{\rm gas}$).  For a given thermal energy, the electrons have a greater mean thermal speed than the ions due to their smaller mass and as a result are far more mobile.  This greater mobility ensures that the electrons are more likely to strike and stick to the grain surface and so the grain quickly becomes negatively charged.  As the net negative charge on the grain grows, further electron attachment is discouraged and is restricted only to those electrons energetic enough to overcome the electrostatic barrier.  Additionally, the plasma ions are attracted to the grain and are accelerated towards it.  The flux of ions are deposited on the grain surface and alter the net charge of the grain, consequently influencing the electron flux towards the grain and further changing the grain's net charge.  The net charge of the dust grain fluctuates in this way until a particle-flux equilibrium is reached where the flux of electrons and ions at the grain surface is equal.  When this is achieved, the surface of the grain is at the floating potential $\phi_{f}$~\cite{bouchoule1999},
\begin{equation}
\phi_{f}=-\frac{k_{B}T_{e}}{2\alpha e},
\end{equation}
where
\begin{equation}
\alpha=\left[\ln{\left(\frac{m_{i}}{2\pi m_{e}}\right)}\right]^{-1}.
\end{equation}
The floating potential is dependent upon the electron plasma temperature $T_{e}$ and the plasma mass ratio  $m_{i}/m_{e}$, where $m_{e}$ and $m_{i}$ are the electronic mass and ionic mass respectively.  Since the natural logarithm in $\alpha$ is largely insensitive to variations in the mass ratio $m_{i}/m_{e}$ we can approximate $\alpha\approx0.1$ without loss of generality.  For increasing electron temperature, the electrons in the bulk plasma become more energetic and so a greater number of electrons have sufficient energy to overcome the electrostatic barrier of a negatively charged grain and reach its surface.  As a result, the grain becomes increasingly negatively charged and the magnitude of its electrostatic potential greater.  

When the floating potential is reached, the dust grain has a constant net negative charge; is surrounded by a plasma sheath (an electron depleted region); and is screened by the bulk plasma.  The sheath length is of the order of the Debye length $\lambda_{D}=(\epsilon_{0}k_{B}T_{e}/(n_{0}e^{2}))^{1/2}$, such that on length scales greater than $\lambda_{D}$ the electric field from the dust grain is zero.  The plasma ions are accelerated from the bulk plasma by the electrostatic field of the grain and are deposited on the grain's surface where they recombine with the surface electrons.  The energy of the plasma ions upon reaching the grain surface may be sufficient to overcome the activation energy of particular chemical reactions that would be unattainable via ion and neutral bombardment from classical, thermal excitation.  As a result, prebiotic molecules or their precursors could be synthesised on the dust grains surface more easily than in the gas-phase.  This is similar to the deposition and manufacture of molecular species on the surface of dust grains in laboratory plasmas~\cite{shi2001,yarin2006,qin2007}.  Note that the energy of the ions may lead to sputtering of the grains mantle, mitigating significant molecular formation.  

The energy gain ($E_{\rm es}$) of an ion carrying charge $q_{i}$ accelerated by the electrostatic field is,
\begin{eqnarray}
E_{\rm es}&=&q_{i}|\phi_{f}| \nonumber \\
&=&\frac{k_{B}T_{e}}{2\alpha},
\end{eqnarray}
where $q_{i}=Z_{i}e$; and $Z_{i}$ is the ion charge number (we will assume that $Z_{i}=1$).  Prior to their energization the ions are in thermal equilibrium with the neutrals ($T_{i}\approx T_{\rm gas}$) and have a thermal of energy of 
\begin{equation}
E_{\rm th}=\frac{3}{2}k_{B}T_{i}.
\end{equation}
The ratio of the ionic electrostatic energy gain relative to the ionic thermal energy is
\begin{equation}
\delta=\frac{E_{\rm es}}{E_{\rm th}}=\frac{T_{e}}{3\alpha T_{i}} \label{delta}.
\end{equation}
When $\delta>1$ the electrostatic energy of the ion is greater than its thermal energy.  Therefore, the total energy of a plasma ion after being accelerated in the electric field of a dust grain is 
\begin{eqnarray}
E_{\rm tot}&=&E_{\rm th}+E_{\rm es} \nonumber \\
&=&(1+\delta)E_{\rm th}.
\end{eqnarray}
In the regime where the electrostatic potential energy of the ion is much greater than its thermal energy, $\delta\gg1$ and the entirety of the ion's energy stems from its electrostatic energization, $E_{\rm tot}\approx \delta E_{\rm th}$.  When $\delta=1$, the electron temperature $T_{e}=3\alpha T_{i}$ and $T_{e}<T_{i}$ inferring that for the electrostatic and thermal energies of the ions to be equal the electrons must be cooler than the ions.  Therefore, in thermal equilibrium the electrostatic energy gain of the ions will always exceed their thermal energy.

\section{Discussion\label{discuss}}
\begin{figure}
\includegraphics{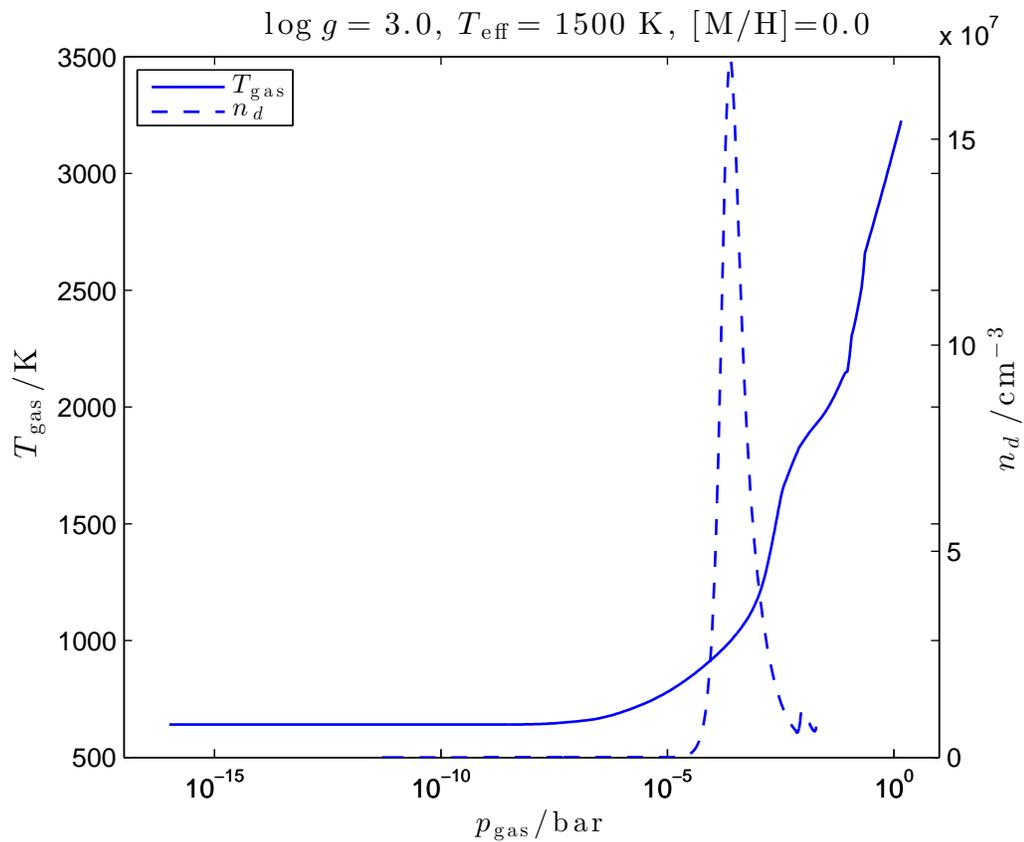}
\caption{$(p_{\rm gas}, T_{\rm gas})$ diagrams and the dust number density $n_{d}$ for an example substeller atmosphere characterized by $\log{g}=3.0$, $T_{\rm eff}=1500$~K and [M/H]=0.0.  These are results from \textsc{Drift-Phoenix} simulations.  \label{fig1}}
\end{figure}

\begin{figure}
\includegraphics{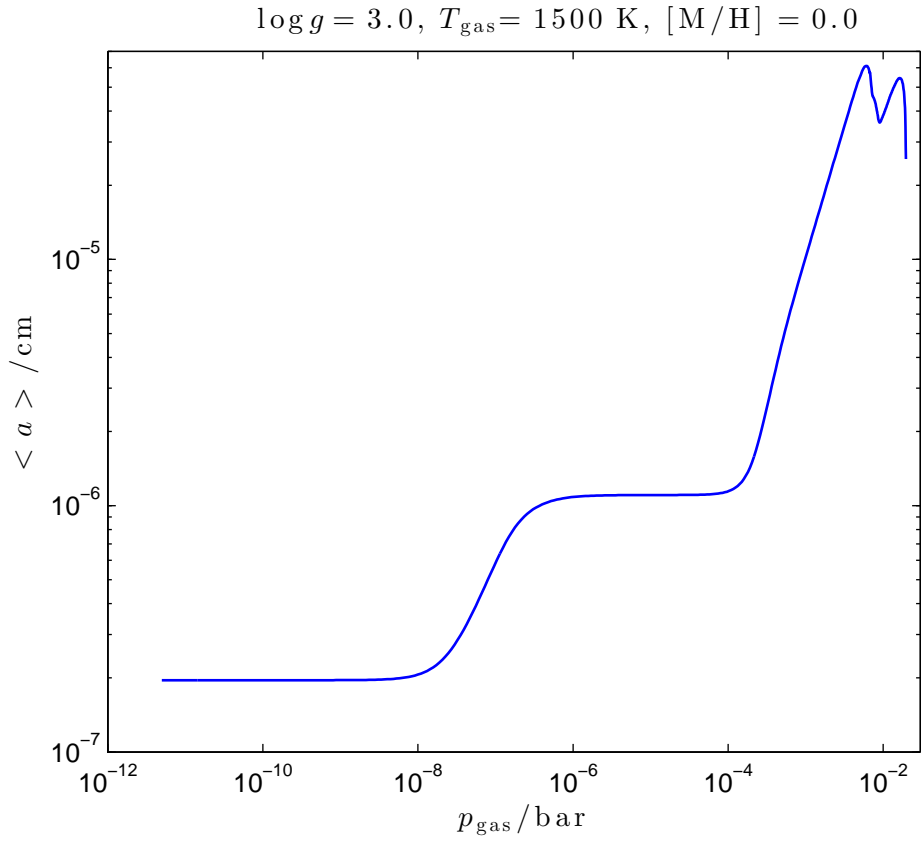}
\caption{Mean grain particle size $\langle a \rangle$ as a function of gas pressure $p_{\rm gas}$ for a exoplanetary atmosphere with $\log{g}=3.0$, $T_{\rm eff}=1500$~K and [M/H]=0.0.\label{fig3}}
\end{figure}

We are interested in such plasma processes in the context of exoplanetary atmospheres (and other substellar objects such as Brown Dwarfs).  We use an example substellar atmosphere using the \textsc{Drift-Phoenix} model atmosphere and cloud formation code~\cite{helling2004,helling2006,helling2008e,hauschildt1999,dehn2007,witte2009,witte2011} defined by $\log{g}=3.0$, $T_{\rm eff}=1500$~K and solar metallicity ([M/H]=0.0).  Figure~\ref{fig1} shows the $(p_{\rm gas}, T_{\rm gas})$ diagram and the dust number density $n_{d}$ of the substellar atmosphere considered here.  In \textsc{Drift-Phoenix} model atmospheres $T_{\rm dust}=T_{\rm gas}$; therefore, for the dust populated regions of the atmosphere the dust temperature is $T_{\rm dust}\approx600-1900$~K.  \textsc{Drift-Phoenix} considers an atmosphere in hydrostatic and chemical equilibrium and uses mixing length theory and radiative transfer theory to consistently calculate the thermodynamic structure of the model atmospheres~\cite{hauschildt1999}.  In addition, \textsc{Drift-Phoenix} kinetically describes the formation of cloud particles as a phase transition process by modelling seed formation, grain growth and evaporation, sedimentation (in phase-non-equilibrium), element depletion and the interaction of these collective processes~\cite{woitke2003,helling2008b,helling2008c}.  Figure~\ref{fig3} shows the mean grain size $\langle a \rangle$ as a function of atmospheric pressure $p_{\rm gas}$.  In the nucleation-dominated upper atmosphere ($p_{\rm gas}\approx10^{-11}$~bar) seed particles form with a mean grain size $\langle a\rangle\approx10^{-7}$~cm.  The dust particles gravitationally settle and grow as they fall, increasing in size.  In the lower atmosphere ($p_{\rm gas}\approx1$~bar) the mean particle size is $\langle a\rangle\approx10^{-5}$~cm.

\begin{figure}
\includegraphics{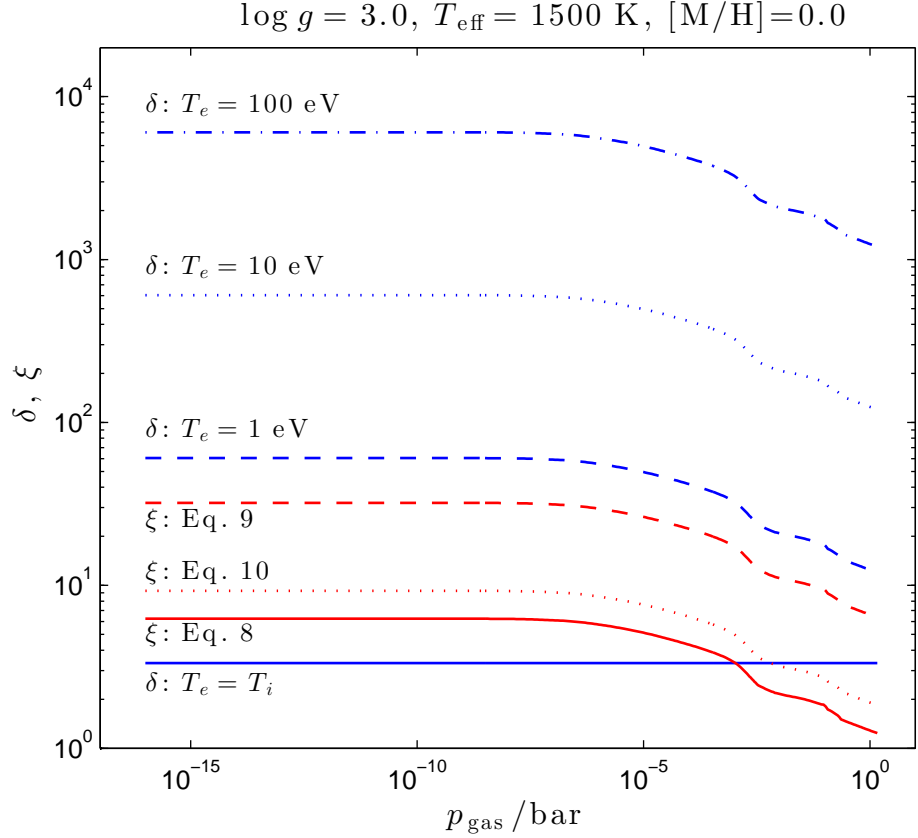}
\caption{$\delta=E_{\rm es}/E_{\rm th}$ and $\xi=E_{0}/E_{\rm th}$ as a function of atmospheric pressure $p_{\rm gas}$ using data from Fig.~\ref{fig1}.  $\delta$ is the ratio of the ionic electrostatic potential energy gain relative to the ionic thermal energy.  $\xi$ is the argument of the Boltzmann factor the chemical reactions considered here:~$\textnormal{CO}+\textnormal{NH}_{3}\rightarrow\textnormal{HCN}+\textnormal{H}_{2}\textnormal{O}$~(Eq.~\ref{rea2}, $E_{0}=0.52$~eV); $\textnormal{CH}_{4}+\textnormal{NH}_{3}\rightarrow\textnormal{HCN}+3\textnormal{H}_{2}$~(Eq.~\ref{rea3}, $E_{0}=2.65$~eV); $\textnormal{CH}_{2}\textnormal{O}+\textnormal{HCN}+\textnormal{NH}_{3}\rightarrow\textnormal{NH}_{3}+\textnormal{NH}_{2}\textnormal{CH}_{2}\textnormal{COOH}$~(Eq.~\ref{rea4}, $E_{0}=0.77$~eV).\label{fig2}}
\end{figure}

Figure~\ref{fig2} shows $\delta$ (Eq.~\ref{delta}, blue lines) as a function of atmospheric pressure $p_{\rm gas}$ for varying electron temperature $T_{e}=1-100$~eV ($\approx10^{4}-10^{6}$~K), for the example atmosphere considered here.  As the electron temperature increases, more electrons are likely to strike and stick to the dust grain and so the local electrostatic field in the plasma sheath is greater in magnitude.  The ions that are accelerated by this electric field gain a greater amount of energy as the electron temperature increases.  

In the scenario where the electrons have had sufficient time to come into thermal equilibrium with the local ions and neutrals ($T_{e}\approx T_{i}\approx T_{\rm gas}$) then $\delta\approx(3\alpha)^{-1}\approx3.33$ (solid blue line).  Even in this case the ions are accelerated by the sheath field to greater energies than that from thermal excitation only.  For example, when $T_{e}\approx T_{i}\approx T_{\rm gas}\approx 600$~K and the atmospheric pressure is $p_{\rm gas}\approx10^{-15}$~bar, the electrostatic energy gain of the ion is $E_{\rm es}\approx0.26$~eV ($\approx 3000$~K); the ionic thermal energy is $E_{\rm th}\approx 0.08$~eV ($\approx 900$~K); and the total energy of the ion upon reaching the grain surface is $E_{\rm tot}\approx0.33$~eV ($\approx 4000$~K).  Consider the surface of a charged dust grain that contains molecules of previously deposited OH.  CH$_{4}$ ions that are accelerated by the sheath electric field from the bulk plasma will strike the surface with an average energy $\approx0.33$~eV, enhancing the likelihood that the chemical reaction $\textnormal{OH}+\textnormal{CH}_{4}\rightarrow \textnormal{CH}_{3}+\textnormal{H}_{2}\textnormal{O}$ (activation energy, $E_{a}=0.2$~eV or $\approx2300$~K) will occur.  If the dust grain populates the atmosphere at an altitude where $p_{\rm gas}\approx10^{-15}$~bar and $T_{i}\approx600$~K, the thermal energy of the ion ($E_{\rm th}\approx 0.08$~eV) would be insufficient to surmount the activation energy for this particular reaction.

At higher electron temperatures, the ions are accelerated to greater energies and so chemical reactions with larger activation energies are accessible.  For example, when $T_{e}=1$~eV (blue dashed line) at $p_{\rm gas}\approx10^{-15}$~bar ($\delta\approx10^{2}$) the total energy gain of the ion is $E_{\rm tot}\approx 4.8$~eV ($\approx9.05\times10^{4}$~K). 

To exemplify the significance of the electrostatic activation of prebiotic chemistry on the surface of a charged dust grain, let's consider the sequence of chemical reactions that lead to the formation of formaldehyde, ammonia, hydrogen cyanide and ultimately the amino acid glycine (NH$_{2}$CH$_{2}$COOH):
\begin{eqnarray}
\textnormal{CH}_{4}+\textnormal{O}_{2}&\rightarrow&\textnormal{CH}_{2}\textnormal{O}+\textnormal{H}_{2}\textnormal{O} \label{rea1}\\
\textnormal{CO}+\textnormal{NH}_{3}&\rightarrow&\textnormal{HCN}+\textnormal{H}_{2}\textnormal{O} ~~~~~~~~~~~~~~~~~~~E_{0}=0.52~\textnormal{eV}\label{rea2}\\
\textnormal{CH}_{4}+\textnormal{NH}_{3}&\rightarrow&\textnormal{HCN}+3\textnormal{H}_{2}~~~~~~~~~~~~~~~~~~~~E_{0}=2.65~\textnormal{eV}\label{rea3} \\
\textnormal{CH}_{2}\textnormal{O}+\textnormal{HCN}+\textnormal{NH}_{3}&\rightarrow&\textnormal{NH}_{3}+\textnormal{NH}_{2}\textnormal{CH}_{2}\textnormal{COOH}~~~~~E_{0}=0.77~\textnormal{eV} \label{rea4}
\end{eqnarray}
We have chosen glycine because it is the simplest of the amino acids.  The reactions~\ref{rea2},~\ref{rea3} and~\ref{rea4} have formation energies (an indicator of the activation energy of the chemical reaction) of $E_{0}\approx0.52$~eV, $2.65$~eV and $0.77$~eV respectively.  Fig.~\ref{fig2} shows $\xi=E_{0}/E_{\rm th}$ as a function of $p_{\rm gas}$ (red lines), where~$E_{\rm th}=\frac{3}{2}k_{B}T_{i}$, $T_{i}\approx T_{\rm gas}$ and using the values for $T_{\rm gas}$ from Fig.~\ref{fig1}.  The plot shows the argument of the Boltzmann factor for the respective chemical reactions and exhibits the energy required for their activation in comparison to $E_{\rm es}$.  For reactions~\ref{rea2} and~\ref{rea3} we assume that one of the reactants has been previously deposited on the grain surface and the other is accelerated from the bulk plasma.  This is reasonable to presume since the multispecies plasma formed in the atmosphere may contain NH$_{3}$, CO and CH$_{4}$ ions.  For reaction~\ref{rea4} the reactants CH$_{2}$O and HCN are presumed to be formed on the grain surface via reactions~\ref{rea1},~\ref{rea2} and~\ref{rea3} and NH$_{3}$ is accelerated from the bulk plasma.  However, note that this reaction is less likely than the preceding bimolecular reactions since it requires the previous reactions to have already occurred.  

High in the atmosphere where $p_{\rm gas}\approx10^{-15}$~bar, the ion temperature is $\approx 600$~K and the resulting thermal energy of an ion is $\approx 0.08$~eV, which is lower than the energies required to form the reaction products above.  However when the electron temperature $T_{e}=1$~eV the ions are accelerated to $E_{\rm tot}\approx 7.8$~eV (blue lines) which surpasses the require formation energies (red lines), increasing the likelihood that these reactions will occur.  At lower atmospheric pressures where it is hotter, the thermal energy can reach $\approx 0.45$~eV ($\approx 5200$~K) and there will exist a population of higher energy ions that, once neutralised on the surface, may be energetic enough to activate the required chemical reactions to form glycine.

The proportion of the grain surface made up of molecular reaction partners depends upon the composition of the ambient plasma and the ionization process that created it.  The ionization processes may be transient and so these conditions could evolve with time enhancing or diminishing the fraction of the grain surface populated by reaction partners.  If the reactant is a minority species then the fraction will be small and the probability of the reaction occurring will be reduced; if the reactant is quite abundant then the probability is enhanced.  This problem is also pertinent for grain surface chemistry where the reactants are acquired from the ambient gas-phase.  However, in contrast to the neutral gas scenario, the enhanced flux of reactants due to the electrostatic acceleration from the bulk plasma enhances the probability of reactions occurring.  In both the plasma and neutral gas cases the problem can be alleviated through hot-atom kinetics where upon being adsorbed onto the grain surface the reactant has relatively high translational energy and diffuses across the surface (e.g. see~\cite{kammler2000}).  This surface trajectory increases the effective fraction of the grain surface covered by the reactant and hence increasing the probability of it encountering and reacting with another species.

Note that in the chemical reactions considered here we assume that the ions and electrons recombine on the surface before their involvement in the neutral-neutral reactions.  Upon recombining the product can either remain on the surface or return to the plasma.  In the former case, particle growth occurs via plasma deposition, this is observed in the laboratory (e.g. see \cite{matsoukas2004}) and so it seems reasonable to presume that the product is retained on the surface.  In the latter case, the ejection of the product will mitigate further surface reactions that the product could participate in and could potentially eject prebiotic and organic molecules into the surrounding gas-plasma mixture.  These molecules may be subsequently reabsorbed by another dust grain.  To ascertain the effect of the grain surface properties (and other factors) on the retention or ejection of the product warrants further investigation.    

Instead of neutral-neutral reactions, the incoming ions may also interact directly with the neutrals via ion-neutral chemistry.  In the ion-neutral case the activation energies are reduced and so easier to occur.  In this scenario the increased flux of ions due to the electrification of the dust grains is what drives the enhancement of chemistry relative to the thermally driven case.  As an example, we list the most likely series of ion-neutral reactions that lead to the formation of glycine~\cite{largo2010}:
\begin{eqnarray}
\textnormal{NH}^{+}_{3}+\textnormal{CH}_{3}\textnormal{COOH}&\rightarrow& \textnormal{NH}^{+}_{4}+\textnormal{CH}_{2}\textnormal{COOH} \\
\textnormal{NH}^{+}_{3}+\textnormal{CH}_{2}\textnormal{COOH}&\rightarrow& \textnormal{NH}_{3}\textnormal{CH}_{2}\textnormal{COOH}^{+} \\
&\rightarrow&\textnormal{NH}_{2}\textnormal{CH}_{2}\textnormal{COOH}^{+}+\textnormal{H}
\end{eqnarray}
The formation of CH$_{3}$COOH (acetic acid) by ion-neutral reactions is not well understood (see also~\cite{blago2003}).  However, it could be synthesised via neutral-neutral reactions on the grain surface in analogy with its formation in interstellar ices containing CO$_{2}$ and CH$_{4}$~\cite{bennett2007}.

The bombardment of the grain surface by electrostatically energised ions also contributes to the heating of the grain surface, increasing the grain temperature $T_{\rm dust}$.  Dust grain heating also occurs due to electron-ion recombination and chemical reactions at the grain surface.  For example, modelling of dust in laboratory plasmas for pressures of $\approx 10^{-3}$ bar and a plasma density of $\approx 10^{17}$~m$^{-3}$, give dust particle temperatures of $\approx 1000$~K~\cite{kilgore1994}.  These calculations considered $0.1$~$\mu$m sized grains composed of aluminium, taking into account heating due to electron-ion recombination, positive ion impact; heat losses due to radiative cooling and Knudsen conduction~\cite{dau1993,kilgore1994}.  Although the situation in substellar atmospheres is different (e.g. grains are composed of multiple materials) these values are indicative as to what to expect in an astrophysical environment where dusty plasmas exist.  In substellar atmospheres the different properties of the dust grown and their radiative behaviour has to be carefully considered~\cite{woitke2003} 

For metal grains, heating can lead to the thermionic emission of electrons from the grain surface, contributing to the charging of the grain.  In \textsc{Drift-Phoenix} model atmospheres the dust is at the same temperature as the gas-phase ($\approx 600-1900$~K, equivalent to $\approx0.05-0.16$~eV).  The work function for most materials is in the range of $\approx 1-7$~eV (or of the order of $10^4$~K).   In general,  \textsc{Drift-Phoenix} dust grains are insulators and so most of the valence electrons are busy participating in bonds, resulting in large work functions $\approx 4-7$~eV (and may even be higher).  Furthermore, for non-conductors there is usually quite a considerable energy band gap between the valence band and the conduction band and so any available electrons find it hard to surmount this barrier to escape.  In the atmospheric model presented here most of the dust grains have a temperature below $0.1$ eV ($\approx10^{3}$~K) and so only work functions $\lesssim 2-3$~eV will be affected by thermionic emission~\cite{wu2005}.  The hottest grains have temperatures $\approx 0.16$~eV ($\approx1900$~K); for these grains thermionic emission starts to become important for work functions $\lesssim 3-4$~eV~\cite{wu2005}.  Therefore, in our model it is assumed that the charging of the dust grain is dominated by the collection of electrons and ions from the surrounding plasma and thermionic emission is a second order effect.

The increase in the thermal energy of the grain can lead to the enhanced thermal activation of surface chemistry provided $k_{B}T_{\rm dust}\geq E_{a}$.  However, ionic bombardment can cause the sputtering of material off the surface of the grain and if the temperature of the grain exceeds a critical value, grain evaporation can occur.  The extent to which this happens is dependent upon the material composition and properties of the dust grain.  Note that grain heating can alter the solid state properties of the grain, potentially changing the internal microstructure of the material resulting in increasing its ductility (cf annealing).

\section{Summary}
Substellar atmospheres are composed of localized volumes of gas-plasma mixtures populated with dust and mineral dust clouds.  In a plasma, dust particles become negatively charged and as a result a plasma sheath forms around the particle accelerating the plasma ions towards the surface.  In this paper we have investigated the energization of plasma ions as they are accelerated from the bulk plasma to the surface of a charged dust grain.  If the ions gain a significant amount of energy, this energy may be sufficient to overcome the activation barrier of certain chemical reactions that would be inaccessible via classical thermal ion and neutral fluxes.  This may lead to an increase in the creation of prebiotic molecules on the grain surface.  The electrostatic potential of the dust grains, and hence the energy gain of the ions upon falling through this potential, is proportional to the plasma electron temperature.  Even in the case where the electrons and ions are in thermal equilibrium ($T_{e}\approx T_{i}$), the ions are accelerated by the sheath field to greater energies than that from thermal excitation only.  To highlight the significance of the electrostatic activation of prebiotic chemistry on the surface of a charged dust grain we considered a simplified sequence of reactions that form formaldehyde, ammonia, hydrogen cyanide and ultimately glycine (a variation of Strecker synthesis).  We found that for modest plasma electron temperatures $T_{e}\approx 1$~eV ($\approx 10^{4}$~K), the energy gain of the accelerated ions is sufficient to exceed the formation energies required for the chemical reactions to occur.  Higher electron temperatures will allow the ions to surmount higher activation barriers and enable other chemical reactions normally inaccessible at low-temperatures.  However, in some scenarios the ionic flux at the grain surface may lead to heating and the ultimate evaporation of the dust grain.

This paper establishes the feasibility of the electrostatic activation of prebiotic chemistry.  This idea can be developed to explicitly model the surface chemical kinetics, describing the incoming accelerated ions interacting with the grain surface.  In this way, the effect of the plasma ionic species, the composition of the grain surface and the effect of the grain charge on the resulting surface chemical reactions can be quantified.

\renewcommand{\abstractname}{Acknowledgements}
\begin{abstract}
The authors would like to thank the anonymous referees for their invaluable comments and suggestions that have helped improved this paper.  ChH, CRS and PBR are grateful for the financial support of the European Community under the FP7 by an ERC starting grant.  DAD is grateful for funding from the UK Science and Technology Funding Council via grant number ST/I001808/1.  We also acknowledge our local IT support.
\end{abstract}

\pagebreak
\bibliographystyle{chicago}

\end{document}